\documentclass[11pt]{article}
\usepackage[preprint]{acl}

\usepackage{xurl}
\usepackage{times}
\usepackage{latexsym}
\usepackage{mathptmx}
\usepackage{amsmath}
\usepackage{amssymb}
\usepackage{array, colortbl} % 用于表格格式与颜色
\usepackage{booktabs}
\PassOptionsToPackage{table}{xcolor}
\usepackage{xcolor}
\usepackage{multirow}        % 合并单元格
\usepackage{graphicx}
\usepackage{bbding} % 加载专用符号包
\usepackage{pifont}
\usepackage{stackengine}
\usepackage[normalem]{ulem}
\usepackage{relsize}
\usepackage{tabularx}
\usepackage{float}
\usepackage{listings}
\usepackage{algorithm}
\usepackage{algorithmicx}
\usepackage{algpseudocode}
\usepackage{enumitem}
\usepackage{siunitx}

\usepackage[T1]{fontenc}
\usepackage[utf8]{inputenc}

\usepackage{microtype}

\usepackage{inconsolata}

\usepackage{graphicx}
\usepackage[affil-it]{authblk}
\definecolor{green1}{HTML}{87CFA4} % 深绿
\definecolor{green2}{HTML}{B8F2A1} % 浅绿
\definecolor{green3}{HTML}{00B200} % 浅绿背景
\definecolor{green4}{HTML}{F0FFF0} % 深绿字体
\definecolor{gray}{HTML}{F4F4F4} % 浅灰背景

\title{Learning to Generate Secure Code via Token-Level Rewards}

\author[1,2]{\textbf{Jiazheng Quan}}
\author[2]{\textbf{Xiaodong Li}}
\author[3]{\textbf{Bin Wang}}
\author[4]{\textbf{Guo An}} 
\author[3]{\textbf{Like Liu}} 
\author[1]{\\\textbf{Degen Huang}}
\author[4,\dag]{\textbf{Liu Lin}}
\author[1,\dag]{\textbf{Chengbin Hou}}

\affil[1]{%
Fuyao University of Science and Technology,
$^2$Xiamen University}
\affil[3]{% 
Peking University,
$^4$Huawei}

\begin{document}

\maketitle
\begin{abstract}

Large language models (LLMs) have demonstrated strong capabilities in code generation, yet they remain prone to producing security vulnerabilities. Existing approaches commonly suffer from two key limitations: the scarcity of high-quality security data and coarse-grained reinforcement learning reward signals. To address these challenges, we propose Vul2Safe, a new secure code generation framework that leverages LLM self-reflection to construct high-confidence repair pairs from real-world vulnerabilities, and further generates diverse implicit prompts to build the PrimeVul+ dataset. Meanwhile, we introduce SRCode, a novel training framework that pioneers the use of token-level rewards in reinforcement learning for code security, which enables the model to continuously attend to and reinforce critical fine-grained security patterns during training. Compared with traditional instance-level reward schemes, our approach allows for more precise optimization of local security implementations. Extensive experiments show that PrimeVul+ and SRCode substantially reduce security vulnerabilities in generated code while improving overall code quality across multiple benchmarks.

% In addition, we introduce SRCode, a training framework that, for the first time, incorporates token-level rewards into reinforcement learning for code security. This design enables the model to continuously attend to and reinforce critical fine-grained security patterns during training. 

\end{abstract}

\section{Introduction}

In recent years, the rapid advancement of large language models (LLMs) has accelerated the adoption of AI-assisted coding tools in industrial settings. These tools are capable of automating software development workflows with expert-level proficiency and generating complex code from natural language prompts, substantially reducing the barrier to software development \cite{hou2024large,du2024evaluating}. However, a growing body of empirical studies and reports reveals a concerning trend: the model-generated code is increasingly being deployed directly into production environments \cite{wang2025large,tabarsi2025llms,park2025continuance}. Despite their strong ability to produce functionally correct implementations, current LLMs remain far from meeting the security requirements necessary for production deployment. Prior work has shown that a substantial fraction of generated code still contains latent security vulnerabilities under realistic evaluations \cite{xu2024large,zhang2025Unseen}. As a result, enhancing the secure code generation capabilities of LLMs has become one of the most urgent challenges in this domain.

% model-generated code is increasingly being directly integrated into real-world engineering projects 

However, the previous work continues to exhibit several limitations when deployed in practical settings \cite{Purpcode,LPO,SafeCoder,Hexacoder}. (i) Existing training data often lacks realism. Many approaches construct datasets using template-based instructions that deviate substantially from the natural contexts in which developers interact with LLMs, making it difficult for fine-tuned models to maintain robust security behavior. (ii) Current training task designs are limited in both hierarchy and diversity. Most datasets consist of isolated, homogeneous tasks without a structured progression of difficulty, which constrains models from developing a principled understanding of security concepts and sustained safety awareness during code generation. 

Beyond the data and task designs, limitations also emerge in how security-related training signals are conveyed during reinforcement learning. (iii) Reinforcement learning reward signals remain insufficiently specified. Although reinforcement learning is widely regarded as a promising mechanism for improving model security, most existing methods rely on coarse, instance-level rewards that provide weak supervision for LLMs. In practice, many security vulnerabilities stem from localized and subtle code patterns, especially in system-level languages such as C/C++. Such coarse reward signals are insufficient for capturing fine-grained safety behaviors, which lead to security degradation when models are deployed in more natural, open-ended, and non-template-based settings.

To enable models to learn fine-grained secure code implementations while reducing the semantic gap between training and real-world deployment, we propose SRCode, a two-stage training framework for secure code generation. Rather than simply aggregating capabilities across isolated tasks, SRCode aims to progressively cultivate security awareness in the training environments that closely resemble real-world usage.

We posit that secure coding ability emerges not from a single task, but through a continuous and semantically coherent training process. To this end, we organize diverse training contexts using multiple implicitly guided prompts and structure three tasks, i.e., vulnerability identification, vulnerability remediation, and secure code generation, into a curriculum with increasing difficulty, resulting in the PrimeVul+ dataset. During the supervised fine-tuning (SFT) stage, the model gradually builds a systematic security knowledge structure through hierarchical and progressively challenging tasks, instead of passively memorizing static secure coding templates from isolated examples. 

Building upon this SFT foundation, the reinforcement learning stage of SRCode further emphasizes the fine-grained structure of secure code. We introduce token-level rewards (TLR) into reinforcement learning for code security, using fine-grained reward signals to explicitly reinforce defensive coding patterns. This design encourages the model to continuously attend to security-relevant tokens during generation, rather than relying on ambiguous guidance from templated instructions or instance-level rewards. Such a fine-grained and curriculum-driven training process not only improves the model’s ability to mitigate localized vulnerabilities, but also enhances its structured understanding of secure implementation strategies, enabling more robust secure code generation in complex and non-template real-world scenarios. 

The contributions are summarized as follows:

\textbf{(1) Realistic, real-world-usage supervision.} We develop the Vul2Safe code generation framework and the PrimeVul+ dataset, which introduce real developer interaction patterns across diverse contexts into training through implicitly guided prompt design and a curriculum of progressively challenging, hierarchical tasks. This design provides supervision for secure code generation that more closely reflects real-world usage scenarios.

% 版本1：To our knowledge, we for the first time introduce token-level rewards (TLR) into reinforcement learning for secure code generation, enabling reinforcement learning to operate at a fine-grained, token-level resolution.

% 版本2：To our knowledge, we are the first to introduce token-level rewards (TLR) into reinforcement learning for secure code generation, enabling reinforcement learning to operate at a fine-grained, token-level resolution.

% 版本3：To our knowledge, we introduce token-level rewards (TLR) into reinforcement learning for the secure code generation setting, enabling reinforcement learning to operate at a fine-grained, token-level resolution.

\textbf{(2) Fine-grained token-level reinforcement.} We introduce token-level rewards (TLR) for secure code generation, enabling reinforcement learning to operate at a fine-grained, token-level resolution for the first time. This design explicitly captures local distinctions between secure and vulnerable patterns, allowing the model to focus on critical details during generation and significantly enhancing the effectiveness of reinforcement learning.

\textbf{(3) Comprehensive empirical evaluation.} We conduct systematic evaluations of SRCode across models with diverse sizes and architectures. Experimental results show consistent and significant improvements across all security-related metrics, as well as stable performance gains under different evaluation settings, highlighting both the effectiveness and robustness of our approach.
% demonstrating the effectiveness and generality of our approach.

%\begin{enumerate}[label=(\arabic*)]
%    \item \textbf{Realistic, real-world-usage supervision.} We develop the Vul2Safe code generation framework and the PrimeVul+ dataset, which introduce real developer interaction patterns across diverse contexts into training through implicitly guided prompt design and a curriculum of progressively challenging, hierarchical tasks. This design provides supervision for secure code generation that more closely reflects real-world usage scenarios.
%    \item \textbf{Fine-grained token-level reinforcement.} We introduce token-level rewards (TLR) for secure code generation, enabling reinforcement learning to operate at a fine-grained, token-level resolution. This design explicitly captures local distinctions between secure and vulnerable patterns, allowing the model to focus on critical details during generation and significantly enhancing the effectiveness of reinforcement learning.
%    \item \textbf{Comprehensive empirical evaluation.} We conduct systematic evaluations of SRCode across models with diverse sizes and architectures. Experimental results show consistent and significant improvements across all security-related metrics, demonstrating the effectiveness and generality of our approach.
%\end{enumerate}

\section{Related Work}

\subsection{Secure Code Generation} 

With the rapid adoption of code-generation LLMs in software development practices \cite{santacoder,codellama,wizardcoder,starcoder,codexglue,qwen2.5-coder,deepseek-coder}, researchers have increasingly recognized their systematic deficiencies in code security. Although existing approaches enable models to generate functionally correct code across diverse tasks \cite{StepCoder,PPOCoder,codeRL}, the generated code often exhibits security flaws in real-world settings, such as missing input validation and improper handling of boundary conditions. These issues largely originate from structural limitations in the training process. 

First, many security-oriented datasets deviate from realistic development contexts, as their instruction designs are typically highly templated and fail to reflect natural developer–model interactions \cite{SVEN,SafeCoder,LPO}. Second, existing training paradigms usually focus on isolated objectives, lacking semantic continuity and hierarchical task progression. Such fragmented supervision makes it difficult for models to form a coherent and systematic understanding of code security, thereby constraining their reasoning capabilities in complex scenarios \cite{wang2023enhancing,LPO}. To mitigate these limitations, several studies have explored security-oriented prompt engineering and preference learning to encourage models to attend to potential vulnerabilities \cite{LPO,Purpcode,Hexacoder}. Nevertheless, these methods struggle to provide sustained and effective guidance in long-sequence generation or scenarios involving complex control flow. 

Moreover, recent analyses \cite{dai2025comprehensive} indicate that training paradigms based on instance-level supervision or sequence-level rewards generally lack fine-grained signals, preventing models from identifying vulnerabilities that are localized to a small number of critical statements.

\subsection{Reinforcement Learning for LLMs} 

Reinforcement learning (RL) has become a key technique in improving LLMs' behavioral consistency and instruction-following. With the development of reinforcement learning from human feedback (RLHF) \cite{RLHF}, various reward- and preference-optimization methods have emerged \cite{RLTF,ProSec}, including policy optimization frameworks based on PPO \cite{PPO}. They allow models to receive continuous feedback during generation, providing a foundation for addressing complex sequential decision problems such as code security. 

However, most safety-oriented code generation approaches still rely on instance- or sequence-level rewards \cite{REAL,Purpcode}, where reward sparsity in long sequences reduces RL’s effectiveness in learning safety semantics. To mitigate this, some studies have explored finer-grained reward mechanisms \cite{CodePRM,LPO,PGRPO,RAMS}, but these either depend on overly complex structured analysis or fail to capture token-level information that leads to vulnerabilities. 

Distinguished from the prior work, we introduce token-level rewards (TLR) into reinforcement learning for code security, enabling fine-grained, security-oriented tuning that directly reinforces token-level safety patterns.

\section{Methodology}

\begin{figure*}[htbp]
    \centering
    \includegraphics[width=0.935\textwidth]{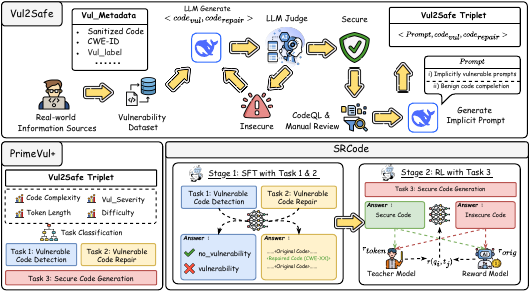}
    \caption{\textbf{Our Methodology.}
    (1) \textbf{Vul2Safe} transforms real-world vulnerable code into high-quality secure repair data with diverse implicit prompts. 
    (2) \textbf{PrimeVul+} ranks the samples based on four metrics and classifies them into three progressively difficult, curriculum-style tasks.
    (3) \textbf{SRCode} first applies SFT on detection and repair tasks, and then performs RL on secure code generation with token-level rewards (TLR) for fine-grained safety optimization.}
    \label{fig:methods}
\end{figure*}

Despite the recent advances in LLM-based code generation, secure code generation remains a significant challenge. To tackle these issues, we propose a systematic solution, as illustrated in Figure \ref{fig:methods}.

\subsection{Vul2Safe: Implicitly Prompted Secure Code Generation} 

To systematically construct high-quality security data for training, we propose the Vul2Safe framework, which converts real vulnerability samples into highly reliable security patch corpora and evaluation tasks. Specifically, we adopt DeepSeek-R1 as the base model and leverage its self-reflection mechanism to perform multi-round generative repairs on vulnerable code. Repair quality is ensured through CodeQL \cite{CodeQL} static analysis and manual sampling audits, producing semantically consistent $<code_{vul},~code_{repair}>$ pairs. 

To better reflect how developers interact with LLMs in natural contexts, Vul2Safe generates diverse prompts for each sample. These prompts are implicitly guided by DeepSeek-R1 to cover different task contexts and usage scenarios. We incorporate two strategies in the implicit prompting: (i) vulnerability-inducing prompts based on functional requirements, which embed unsafe implementation patterns in instructions to guide the model toward vulnerability-related contexts; (ii) code-completion prompts based on benign code prefixes, which present potentially risky snippets to elicit the model’s understanding of vulnerability patterns during subsequent code generation. Further details can be found in Appendix \ref{appendixH}. Ultimately, Vul2Safe produces $<Prompt,~code_{vul},~code_{repair}>$ triplets, automatically transforming vulnerable code into structurally standardized, task-reproducible secure repair corpora, effectively addressing the scarcity of high-quality security data.

\subsection{PrimeVul+: Curriculum-Based Safe Code Generation Dataset} 

% Guided by 

To support training and evaluation of LLMs in real-world code security scenarios, we construct the PrimeVul+ dataset from high-quality open-source code. 
Following OWASP's latest LLM security report \cite{owasp-llm-top10-2025}, we select 3,289 code samples from PrimeVul \cite{PrimeVul} covering 14 common weakness enumeration (CWE) vulnerability categories. Each sample is systematically processed using the Vul2Safe framework. After that, we evaluate and organize the tasks according to vulnerability complexity and remediation difficulty, resulting in PrimeVul+, a structured and quality-controlled dataset closely aligned with real-world LLM usage. The dataset includes three progressively challenging task categories: vulnerable code detection, vulnerable code repair, and secure code generation, totaling 2,500 tasks. 

Concretely, the first two categories are used for SFT to improve the model's basic capabilities in identifying and fixing vulnerabilities. The third category, i.e., secure code generation, is designed for RL, training the model to generate robust, best-practice-compliant security implementations under complex security contexts. PrimeVul+ thus provides a high-quality, scalable, and realistic foundation for training, evaluating, and benchmarking code security models.

\subsection{TLR: Token-level Rewards for Safe Code Generation}
\label{sec:TLR}

In reinforcement learning (RL) driven secure code generation, traditional reward signals are typically evaluated at the code fragment level, which makes it difficult for policy models to accurately learn local secure implementations. To address this, we introduce a token-level rewards (TLR) mechanism that assigns fine-grained rewards to each token generated by the model. Specifically, for each training sample, the policy model receives an input prompt $q_i$ and generates a code sequence $S_i = (t_0, t_1, ..., t_{L_i - 1})$ of length $L_i$, token by token. After generating the sequence, the teacher model identifies secure implementations within the code and extracts all corresponding tokens into the secure token set $\textit{T}$. Token-level positive rewards $r_{\text{token}}^+(q_i, t_j)$, for $j \in [0, L_i - 1]$, are then computed based on $\textit{T}$, as defined in Equation \eqref{eq1}.
\begin{equation}
    \label{eq1}
    r_{\text{token}}^+(q_i, t_{j}) = 
        \begin{cases} 
        0, & t_{j} \notin \textit{T} \\
        \alpha, & t_{j} \in \textit{T}
        \end{cases}
\end{equation}
where $\alpha > 0$ represents the fixed reward for security tokens. The teacher model then analyzes whether the code contains CWE vulnerabilities. If they exist, $\text{Vul}(S_i) = True$ is set; a negative reward $r_{global}$ is assigned based on the number and severity of vulnerabilities. This yields the negative reward $r_{\text{token}}^-(q_i, t_{j})$, as depicted in Equation \eqref{eq2}.
\begin{equation}
    \label{eq2}
    r_{\text{token}}^-(q_i, t_{j}) = 
        \begin{cases} 
        0, & \neg\text{Vul}(S_i) \\
        \frac{r_{\text{global}}}{L_i}, & \text{Vul}(S_i)
        \end{cases}
\end{equation}

By combining the above process, the instantaneous reward $r(q_i, t_{j})$ is obtained for each token, as shown in Equation \eqref{eq3}, where $r_{\text{orig}}(q_i, t_{j})$ represents the reward generated by reward model. 
\begin{equation}
    \label{eq3}
    \begin{aligned}
        r(q_i, t_{j}) & = r_{\text{orig}}(q_i , t_{j}) + r_{\text{token}}^-(q_i, t_{j}) \\
        & + r_{\text{token}}^+(q_i, t_{j})
    \end{aligned}
\end{equation}

Regarding RL policy optimization, the gradient follows the standard REINFORCE form \cite{REINFORCE}, as described in Equation \eqref{eq4} below:
\begin{equation}
    \label{eq4}
    \nabla_\theta J(\theta) = \mathbb{E}\left[ \sum_j \nabla_\theta \log \pi_\theta(t_j \mid t_{<j}) A_j \right]
\end{equation}

The advantages and further details (e.g., pseudo-code) of TLR are elaborated in Appendix \ref{appendixA}.

\subsection{SRCode: A Two-Stage Training Framework Using the TLR}

In this work, we propose SRCode, a two-stage training framework incorporating the TLR mechanism. During the SFT stage, we fine-tune baseline models on the vulnerable code detection and vulnerable code repair tasks using the PrimeVul+ dataset, thereby enhancing their fundamental capabilities to detect and repair vulnerabilities. During the RL stage, the policy model generates each token in the code sequence to obtain immediate rewards as defined by Equation \eqref{eq3}.

Upon receiving immediate rewards, the generalized advantage estimation (GAE) is employed to construct a token-level advantage function. To this end, we define the state during generation process as a prefix representation $s_{i, j} = (q_i, t_{<j})$ and denote the value network as $V_\phi(\cdot)$. The immediate reward for generating token $t_j$ under state $s_{i, j}$ is denoted as $r_{i, j} = r(q_i, t_j)$. Based on above definitions, the one-step temporal difference residual becomes as:
\begin{equation}
    \label{eq5}
    \delta_{i,j} = r_{i,j} + \gamma V(s_{i,j+1}) - V(s_{i,j})
\end{equation}
where $\gamma \in [0, 1]$ is the discount factor. After that, we employ GAE to perform weighted accumulation of residuals from future time steps, thus obtaining the advantage estimate $A_{\text{secure}}(q_i, t_j)$ for token $t_j$:
\begin{equation}
    \label{eq6}
    A_{\text{secure}}(q_i, t_j) = \sum_{l=0}^{L_i - j - 1} (\gamma \lambda)^l \delta_{i,j+l}
\end{equation}
where $\lambda \in [0, 1]$ controls the balance between bias and variance. At sequence termination states $s_{i, L_i}$, we set $V_\phi(s_{i, L_i}) = 0$ to ensure the advantage estimate converges naturally at the end. This definition provides a stable and variance-controlled estimation method for the advantage term in the subsequent PPO objective function, fully compatible with our introduced token-level fine-grained reward signal mechanism.

Based on this, we use the aforementioned advantage function $A_{\text{secure}}(q_i, t_j)$ as the core input to the PPO clip objective, with the basic form of the objective function given by Equation \eqref{eq7}.

\begin{small}
\begin{equation}
\setlength{\abovedisplayskip}{-2pt}
    \label{eq7}
    \begin{aligned}
        J_{\text{PPO}}^{\text{secure}} &= \sum_{i=1}^M \sum_{j=0}^{L_i - 1} \min\left( \rho_{i,j}\times A_{\text{secure}}(q_i, t_j), \right. \\
        &\quad \left. \text{clip}\left( \rho_{i,j}, 1 - \epsilon, 1 + \epsilon \right) \times A_{\text{secure}}(q_i, t_j) \right)
    \end{aligned}
\end{equation}
\end{small}
where $\rho_{i, j}$ represents the probability ratio between the current policy and the old policy at token $t_j$, which corresponds to the importance sampling coefficient in the PPO framework. To prevent excessive updates, we apply clipping to this coefficient. The specific formula for the importance sampling coefficient is as follows:
\begin{equation}
    \label{eq8}
    \rho_{i,j} = \frac{P_\theta(t_j \mid s_{i,j})}{P_{\theta'}(t_j \mid s_{i,j})}, \quad s_{i,j} = (q_i, t_{<j})
\end{equation}

In the actual optimization process, the PPO loss function takes the negative value of the aforementioned objective function. The gradient formula \eqref{eq9} demonstrates that the log probability of each token is multiplied by the pruned advantage function, thereby reinforcing safe tokens and suppressing vulnerability patterns during updates.

\begin{small}
\begin{equation}
\setlength{\abovedisplayskip}{-2pt}
    \label{eq9}
    \begin{aligned}
        \nabla_{\theta} \mathcal{L}_{\text{PPO}}^{\text{secure}} &= -\sum_{i=1}^M \sum_{j=0}^{L_i - 1} \bigg[ \min\bigg( \rho_{i,j}\times A_{\text{secure}}(q_i, t_j), \\
        &\quad \text{clip}(\rho_{i,j}, 1 - \epsilon, 1 + \epsilon)\times A_{\text{secure}}(q_i, t_j) \bigg) \\
        &\quad \times \nabla_{\theta} \log P_{\theta}(t_j \mid s_{i,j}) \bigg]
    \end{aligned}
\end{equation}
\end{small}

\section{Experiments}

To evaluate the effectiveness of SRCode in enhancing code generation security and study its underlying mechanisms, we group experiments and pose the following four research questions (RQs). 

% These questions comprehensively examine the efficacy and potential limitations of our approach from complementary perspectives: overall method performance, reliability of security improvements, sampling efficiency, and generalizability.

\textbf{RQ1: Effectiveness Improvement Analysis.} \textit{Compared to baseline methods, do our PrimeVul+ dataset and SRCode approach substantially enhance the safety of code generated by LLMs?}

\textbf{RQ2: Reliability of Safety Improvements.} \textit{Can SRCode consistently deliver stable performance across models of varying scales and architectures, while uniformly reducing the number of vulnerabilities in the generated code?} 

\textbf{RQ3: Module Contribution Analysis.} \textit{What contribution does each module of SRCode make? Which designs are most critical to improve safety?}

\textbf{RQ4: Generalization Analysis and Sampling Efficiency.} \textit{How does SRCode affect general code generation performance and sampling efficiency?}

% How does SRCode affect general code generation performance and sampling efficiency relative to existing methods?

\begin{table*}[!ht]
  \centering
  \resizebox{\textwidth}{!}{
  {\fontsize{11pt}{13pt}\selectfont
  \renewcommand{\arraystretch}{1.0} 
  \begin{tabular}{l c c c c c}
    \toprule[1.5pt]
    \multicolumn{1}{c}{\multirow{2}{*}{\textbf{Method}}} & \multirow{2}{*}{\textbf{Size}} & \multicolumn{2}{c}{\textbf{CWEval}} & \textbf{CodeLMSec} & \textbf{CyberSecEval} \\
    \cmidrule(lr){3-4}\cmidrule(lr){5-5}\cmidrule(lr){6-6}
    & & FS@1(C/C++) & FS@1(Python) & Sec.Rate & Pass.Rate \\
    \midrule
    \rowcolor{gray} \multicolumn{6}{c}{w/o RL Models} \\
    \textit{Qwen series models} & & & & & \\
    \quad \quad Qwen2.5-Coder-Instruct & 7B & 28.9 & 45.0 & 75.9 & 61.9 \\
    \quad \quad Qwen2.5-Coder-Instruct [SafeCoder*] & 7B & 19.2 & 43.5 & 87.2 & 70.8 \\
    \rowcolor{green4} \quad \quad Qwen2.5-Coder-Instruct [PrimeVul+, ours] & 7B & $\underline{32.7}^{\textcolor{green3}{\uparrow\,3.8\%}}$ & $47.8^{\textcolor{green3}{\uparrow\,2.8\%}}$ & $86.0^{\textcolor{green3}{\uparrow\,10.1\%}}$ & $70.3^{\textcolor{green3}{\uparrow\,8.4\%}}$ \\
    \textit{DeepSeek series models} & & & & & \\
    \quad \quad DeepSeek-Coder-Instruct & 6.7B & 25.0 & 32.0 & 57.4 & 57.9 \\
    \quad \quad DeepSeek-Coder-Instruct [SafeCoder*] & 6.7B & 23.1 & 24.0 & 60.4 & 72.3 \\
    \rowcolor{green4} \quad \quad DeepSeek-Coder-Instruct [PrimeVul+, ours] & 6.7B & $30.8^{\textcolor{green3}{\uparrow\,5.8\%}}$ & $34.9^{\textcolor{green3}{\uparrow\,2.9\%}}$ & $82.5^{\textcolor{green3}{\uparrow\,25.1\%}}$ & $62.9^{\textcolor{green3}{\uparrow\,5.0\%}}$ \\
    \textit{Codellama series models} & & & & & \\
    \quad \quad Codellama-Instruct-hf & 7B & 13.5 & 21.7 & 67.2 & 62.9 \\
    \quad \quad Codellama-hf [SafeCoder] & 7B & 8.4 & 11.8 & 80.5 & 72.3 \\
    \rowcolor{green4} \quad \quad Codellama-Instruct-hf [PrimeVul+, ours] & 7B & $26.9^{\textcolor{green3}{\uparrow\,13.4\%}}$ & $30.4^{\textcolor{green3}{\uparrow\,8.7\%}}$ & $83.0^{\textcolor{green3}{\uparrow\,15.8\%}}$ & $\underline{73.8}^{\textcolor{green3}{\uparrow\,10.9\%}}$ \\
    \midrule
    \rowcolor{gray} \multicolumn{6}{c}{RL Models (Using Qwen2.5-Coder-Instruct as the backbone)} \\
    \multicolumn{1}{c}{Vanilla PPO} & 7B & 26.9 & 45.8 & 84.5 & 65.8 \\
    \multicolumn{1}{c}{Purpcode} & 14B & 25.0 & \uline{52.0} & \textbf{89.2} & 69.8 \\
    \rowcolor{green4} \multicolumn{1}{c}{SRCode (ours)} & 7B & $\textbf{38.5}^{\textcolor{green3}{\uparrow\,9.6\%}}$ & $\textbf{54.2}^{\textcolor{green3}{\uparrow\,9.2\%}}$ & $\underline{88.7}^{\textcolor{green3}{\uparrow\,12.8\%}}$ & $\textbf{74.8}^{\textcolor{green3}{\uparrow\,12.9\%}}$ \\
    \bottomrule[1.5pt]
  \end{tabular}
  }}
  \caption{\fontsize{9bp}{10bp}\textbf{Main Results.} We present the results under the experimental setup described in Section 4.1. \textbf{FS@1}, i.e., func-sec@1, evaluates both the correctness and security of the code; \textbf{Sec.Rate} and \textbf{Pass.Rate} both represent the proportion of securely implemented code out of the total samples; the higher is better.  [PrimeVul+] indicates the model trained using our dataset; [SafeCoder] indicates the model provided by the authors; [SafeCoder*] indicates the model trained by us using the authors' dataset and setting. We highlight the \textbf{best} and \underline{second-best} results, and calculate the improvements of our method compared to the baselines without security optimizations.}
  \label{tab:model_security_perf}
\end{table*}

\subsection{Experienmental Setup}

\textbf{Base Models and Benchmarks.} Three open-source large language series models: CodeLlama \cite{codellama}, Qwen-Coder \cite{qwen2.5-coder}, and DeepSeek-Coder \cite{deepseek-coder} are adopted for experiments. During evaluation, we mainly focus on the effectiveness of LLMs in enhancing code safety within C/C++ scenarios. Specifically, we employ three high-quality security evaluation benchmarks, i.e., CWEval \cite{CWEval}, CodeLMSec  \cite{CodeLMSec}, and CyberSecEval \cite{CyberSecEval}, to ensure our results accurately and comprehensively reflect the models' security capabilities. Besides, to further gauge the model's general code generation capabilities, we employ the HumanEval Pro and MBPP Pro benchmarks \cite{HumanEvalPro}. Detailed experimental settings are provided in Appendix \ref{appendixC}.

\textbf{Baselines.} To systematically evaluate the performance of proposed SRCode framework, we select the representative open-source instruction-fine-tuned large models, including Qwen2.5-Coder-Instruct, DeepSeek-Coder-Instruct, and CodeLlama-Instruct. In comparative experiments, we further introduce SafeCoder \cite{SafeCoder} and DiSCo \cite{LPO} as additional references. Regarding RL optimization methods, we choose the classic PPO approach and the recent high-quality work Purpcode \cite{Purpcode} as baselines. These were compared with the proposed SRCode to verify whether it outperforms existing security fine-tuning methods.
% These models possess only general-purpose code generation capabilities and have not undergone specialized security optimization, serving as foundational references for evaluating security enhancement methods.

\textbf{Evaluation.} We comprehensively evaluate the model-generated code via a combination of static analysis, execution testing, and general-purpose benchmarks to ensure the reliability of our analysis. For pass@1, we report the results using 64 independent samples. 
Further details of benchmarks and metrics can be found in Appendix \ref{appendixB}.

% Regarding benchmark configuration, we fully adopt the official evaluation metrics and validators. Additionally, to measure the impact of our approach on the model's programming capabilities, we use the standard pass@1 metric to assess the model's general code generation ability on the HumanEval Pro and MBPP Pro benchmarks. 

\subsection{RQ1: Effectiveness Improvement Analysis}

To address RQ1, we conduct a systematic evaluation of various models under identical experimental conditions. The evaluation aims to quantify the direct security benefits of the proposed SRCode method while simultaneously examining its potential impact on code functional correctness. Overall, the results, as shown in Table \ref{tab:model_security_perf}, clearly demonstrate that SRCode exhibits significant advantages across all key security metrics.

In particular, these results empirically validate the effectiveness and generalizability of the proposed PrimeVul+ dataset in guiding models to learn secure coding behaviors (see the upper part of Table \ref{tab:model_security_perf}). Specifically, without RL, significant variations in security metrics are observed across different architectures and model families. However, after fine-tuning on PrimeVul+, all models achieve consistent and substantial gains in code security, with an average increase of 9.39\%. Among them, the Codellama model achieves the most pronounced average improvement of 12.20\%, while the DeepSeek model demonstrates an even higher improvement of 25.1\% on CodeLMSec.

% Among baseline models without RL, significant variations in security metrics are observed across different models. However, after fine-tuning on the PrimeVul+ dataset, all models achieve consistent and substantial improvements in code security, with an average increase of 9.39\%. Among them, the Codellama model achieves the most pronounced average improvement of 12.20\%; the DeepSeek model demonstrates an even higher improvement of 25.1\% on CodeLMSec. These results empirically validate the effectiveness and generalizability of the PrimeVul+ dataset in guiding models to learn secure behaviors.

% RL methods demonstrate

After introducing RL, the methods with RL demonstrate an overall positive impact on the safety performance of baseline models (see the lower part of Table \ref{tab:model_security_perf}). However, traditional PPO exhibits a 2.0\% decline in the FS@1 metric compared to the baseline, indicating that without explicit safety objective constraints, traditional PPO tends to prioritize optimizing the overall reward signal. This tendency leads to capability degradation in complex safety scenarios. The further comparison with Purpcode reveals that its GRPO-based optimization strategy achieves commendable results across multiple metrics. Nevertheless, even against this highly competitive baseline model, SRCode demonstrates sustained and significant advantages. Notably, the 7B model trained with SRCode even outperforms the Purpcode's 14B model on nearly all security metrics. This observation implies that the performance gains primarily stem from the method's effective guidance of the model, while also enabling the model to learn security coding conventions with strong transferability.

\subsection{RQ2: Reliability of Safety Improvements}

In security code generation tasks, the focus is not on the model's rote memorization of specific vulnerability patterns, but rather on its ability to learn security coding behaviors with generalization capabilities. Consequently, we analyze and discuss RQ2. Under a unified training setup, we evaluate multiple models spanning different architectures and scales, with the results reported in Table \ref{tab:compare}.

% Under a unified training setup, we evaluate models of varying scales and architectures. The experimental results are displayed in Table \ref{tab:compare}.

\begin{table}[!ht]
\centering
\resizebox{0.5\textwidth}{!}{
\begin{tabular}{>{\centering\arraybackslash}m{3cm} c c c}
\toprule[1.5pt]
\multicolumn{1}{c}{\multirow{2}{*}{\textbf{Method}}}
& \multicolumn{1}{c}{\textbf{CWEval}} 
& \multicolumn{1}{c}{\textbf{CodeLMSec}}
& \multicolumn{1}{c}{\textbf{CyberSecEval}}
\\
\cmidrule(lr){2-2}\cmidrule(lr){3-3}\cmidrule(lr){4-4}
& FS@1(C/C++) & Sec.Rate & Pass.Rate \\
\midrule

\multicolumn{4}{l}{\textit{Qwen2.5-Coder-3B-Instruct}} \\
\cmidrule(lr){1-4}
Base LLM  & \larger[1.5]{21.2} & \larger[1.5]{73.4} & \larger[1.5]{60.9} \\
DiSCo     & \larger[1.5]{15.4} & \larger[1.5]{76.4} & \larger[1.5]{58.4} \\
SafeCoder & \larger[1.5]{13.5} & \larger[1.5]{76.7} & \larger[1.5]{\textbf{65.8}} \\
\rowcolor{gray}
SRCode (\textbf{Ours}) & \larger[1.5]{\textbf{23.1}} & \larger[1.5]{\textbf{79.0}} & \larger[1.5]{65.4} \\

\multicolumn{4}{l}{
\rule{0pt}{3ex} 
\textit{Qwen2.5-Coder-7B-Instruct}} \\
\cmidrule(lr){1-4}
Base LLM   & \larger[1.5]{28.9} & \larger[1.5]{75.9} & \larger[1.5]{61.9} \\
DiSCo      & \larger[1.5]{23.1} & \larger[1.5]{80.0} & \larger[1.5]{57.9} \\
SafeCoder  & \larger[1.5]{19.2} & \larger[1.5]{87.2} & \larger[1.5]{70.8} \\
\rowcolor{gray}
SRCode (\textbf{Ours}) & \larger[1.5]{\textbf{38.5}} & \larger[1.5]{\textbf{88.7}} & \larger[1.5]{\textbf{74.8}} \\

\multicolumn{4}{l}{
\rule{0pt}{3ex} 
\textit{DeepSeek-Coder-6.7B-Instruct}} \\
\cmidrule(lr){1-4}
Base LLM   & \larger[1.5]{25.0} & \larger[1.5]{57.4} & \larger[1.5]{57.9} \\
DiSCo      & \larger[1.5]{21.2} & \larger[1.5]{84.0} & \larger[1.5]{57.4} \\
SafeCoder  & \larger[1.5]{23.1} & \larger[1.5]{60.4} & \larger[1.5]{72.3} \\
\rowcolor{gray}
SRCode (\textbf{Ours}) & \larger[1.5]{\textbf{32.7}} & \larger[1.5]{\textbf{86.7}} & \larger[1.5]{\textbf{73.8}} \\
\bottomrule[1.5pt]
\end{tabular}
}
\caption{\textbf{Comparison of security-enhanced code generation methods.} The best results are in bold.}
\label{tab:compare}
\end{table}
% Comparison of different methods under the same settings.

The overall results show that existing security enhancement methods fail to deliver the expected security improvements in certain scenarios, even exhibiting performance degradation across multiple benchmarks. Despite incorporating substantial security examples during training, these approaches fundamentally rely on pattern-level memorization rather than semantic-level reasoning. In contrast, SRCode achieves stable and significant improvements across all models and benchmarks. 

Taking the Qwen2.5-Coder-7B-Instruct as an example, it achieves improvements of 9.6\%, 12.8\%, and 12.9\% across three metrics, representing the best results within its group. This further demonstrates that the SRCode training strategy enhances transferable and generalizable structured security reasoning capabilities, whose effectiveness is independent of specific model scale or architecture. 

The in-depth analysis of vulnerability counts at different severity levels (high, medium, low) is provided in Appendix \ref{appendixE}, which confirms that SRCode delivers stable and reliable safety improvements across models of varying scales and architectures.

\subsection{RQ3: Module Contribution Analysis}

Table \ref{tab:ablation study} presents the ablation study results. Overall, removing each module individually leads to varying degrees of performance degradation in SRCode. First, the model without SFT shows a $4 \sim 5\%$ decline across all metrics, particularly on tasks involving multi-step pointer operations, memory copying, and boundary-sensitive logic. Without this stage, models trained solely through RL generated code that is more prone to vulnerabilities in CWEval dynamic testing. This indicates that lacking SFT weakens the model's foundational understanding of secure coding conventions and significantly impacts the effectiveness of RL training.

\begin{table}[htbp]
  \centering
  \resizebox{0.5\textwidth}{!}{
  \begin{tabular}{@{}ccc>{\centering\arraybackslash}p{3cm}cc@{}}
    \toprule[1.5pt]
    \multicolumn{3}{c}{\textbf{Method}} & \textbf{CWEval} & \textbf{CodeLMSec} & \textbf{CyberSecEval} \\
    \cmidrule(lr){1-3}\cmidrule(lr){4-4}\cmidrule(lr){5-5}\cmidrule(lr){6-6}
    SFT & RL & TLR & FS@1 & Sec.Rate & Pass.Rate \\
    \midrule[1.2pt]
    {\larger[1.5]{\ding{52}}} & \larger[1.5]{\ding{52}} & \larger[1.5]{\ding{52}} & \textbf{\larger[1.5]{38.5}} & \textbf{\larger[1.5]{88.7}} & \textbf{\larger[1.5]{74.8}} \\
    \larger[1.5]{\ding{55}} & \larger[1.5]{\ding{52}} & \larger[1.5]{\ding{52}} & \larger[1.5]{34.6}  & \larger[1.5]{86.7}  & \larger[1.5]{67.3}  \\
    \larger[1.5]{\ding{52}} & \larger[1.5]{\ding{55}} & \larger[1.5]{\ding{55}} & \larger[1.5]{32.7}  & \larger[1.5]{85.9}  & \larger[1.5]{68.8}  \\
    \larger[1.5]{\ding{52}} & \larger[1.5]{\ding{52}} & \larger[1.5]{\ding{55}} & \larger[1.5]{30.8}  & \larger[1.5]{83.9}  & \larger[1.5]{66.8}  \\
    \larger[1.5]{\ding{55}} & \larger[1.5]{\ding{52}} & \larger[1.5]{\ding{55}} & \larger[1.5]{26.9}  & \larger[1.5]{84.5}  & \larger[1.5]{65.8}  \\
    \larger[1.5]{\ding{55}} & \larger[1.5]{\ding{55}} & \larger[1.5]{\ding{55}} & \larger[1.5]{28.9}  & \larger[1.5]{75.9}  & \larger[1.5]{61.9}  \\
    \bottomrule[1.5pt]
  \end{tabular}
  }
  \caption{\textbf{Ablation study} on Qwen2.5-Coder-7B-Instruct. SFT refers to supervised fine-tuning on the PrimeVul+, RL refers to reinforcement learning using PPO, and TLR refers to the token-level rewards.}
  \label{tab:ablation study}
\end{table}

\begin{figure*}[!ht]
    \centering    \includegraphics[width=0.98\textwidth]{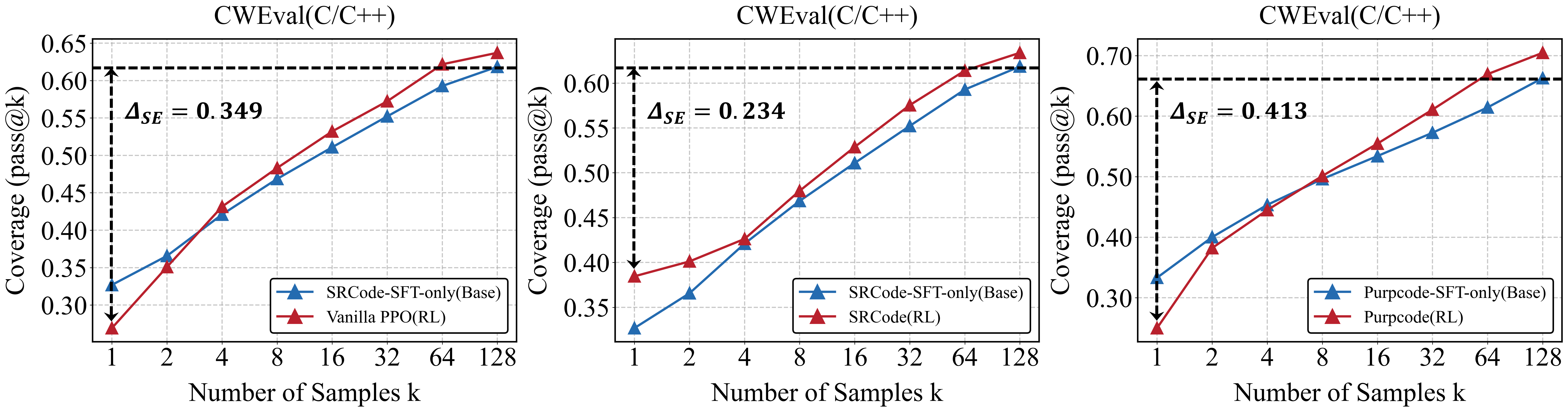}
    \caption{\textbf{Comparison of Sampling Efficiency Gaps ($\Delta_{SE}$).} A smaller $\Delta_{SE}$ indicates that the RL-trained model is closer to the theoretical performance upper bound of the base model under large-k sampling.}
    \label{fig:deltaSE}
\end{figure*}
% Lower sampling counts better reflect real user experience. To enable a precise comparison of RL-induced improvements, all base models are uniformly set to their SFT-only versions.

% upon removing the RL phase, the 
% —exhibiting a classic alignment tax phenomenon
Second, when the RL phase is removed, the FS@1 metric of models shows a significant decline, with the synergistic performance of functionality and security markedly deteriorating, reflecting a classic alignment tax phenomenon. Performance also drops by 2.8\% and 6.0\% on CodeLMSec and CyberSecEval, respectively. The model first learns fundamental functional correctness on the SFT foundation while acquiring preliminary code security implementation. Crucially, it should progressively learn vulnerability-fixing strategies during the RL phase to achieve a reliable balance between functional correctness and security. 

Finally, with TLR removed, all metrics exhibit a clear decline, particularly in the model's fine-grained performance for complex security scenarios. This degradation is most pronounced in tasks involving ambiguous trigger conditions and vulnerabilities caused by the single-character errors. Appendix \ref{appendixG} further presents a case study to intuitively illustrate the advantages of TLR. Besides, more detailed analysis and the ablation results on PrimeVul+ dataset are provided in Appendix \ref{appendixD}.

\subsection{RQ4: Generalization Analysis and Sampling Efficiency}
To verify whether improving code security comes at the cost of utility, we evaluate SRCode over HumanEval Pro and MBPP Pro, two more comprehensive datasets for testing general code generation capabilities with self-calling and code reuse requirements, as shown in Table \ref{tab:code_model_perf}.

% To verify the generalization capability (...), we evaluate SRCode on HumanEval Pro and MBPP Pro, two more comprehensive datasets (...) for testing general code generation capabilities as described in Appendix \ref{appendixB}. 

% To address RQ4, we comprehensively evaluate the performance of different reinforcement learning methods on secure code generation tasks. Specifically, we evaluate the impact of SRCode on the general code generation capabilities of models, with results shown in Table \ref{tab:code_model_perf}. 
\begin{table}[htbp]
    \centering
    \resizebox{0.5\textwidth}{!}{
    \renewcommand{\arraystretch}{1.2}
    \begin{tabular}{cll}
        \toprule[1.5pt]
        \multicolumn{1}{c}{\textbf{Model}} & \multicolumn{1}{c}{\textbf{HumanEval Pro}} & \multicolumn{1}{c}{\textbf{MBPP Pro}} \\
        \midrule
        Qwen2.5-Coder-3B-Instruct & \larger[1.5]{59.1} & \larger[1.5]{55.0} \\
        \quad \textit{w/ SRCode}  & \larger[1.5]{$60.4^{\textcolor{green3}{\uparrow\,1.3\%}}$} & \larger[1.5]{$55.6^{\textcolor{green3}{\uparrow\,0.6\%}}$} \\
        Qwen2.5-Coder-7B-Instruct & \larger[1.5]{65.9} & \larger[1.5]{64.8} \\
        \quad \textit{w/ SRCode}  & \larger[1.5]{$69.5^{\textcolor{green3}{\uparrow\,3.6\%}}$} & \larger[1.5]{64.8} \\
        DeepSeek-Coder-6.7B-Instruct & \larger[1.5]{55.6} & \larger[1.5]{57.1} \\
        \quad \textit{w/ SRCode}  & \larger[1.5]{55.6} & \larger[1.5]{$57.9^{\textcolor{green3}{\uparrow\,0.8\%}}$} \\
        \bottomrule[1.5pt]
    \end{tabular}
    }
    \caption{Evaluation of \textbf{general code generation capability} for the models with SRCode (i.e., w/SRCode).}
    \label{tab:code_model_perf}
\end{table}

It can be observed from Table \ref{tab:code_model_perf} that across three base models of varying scales and architectures, SRCode does not cause any performance degradation, maintaining stable overall performance with the improvements in some cases. Concretely, Qwen2.5-Coder-7B-Instruct achieves a significant 3.6\% increase on HumanEval Pro, while others also show modest gains across various benchmarks. This demonstrates that SRCode enhances model safety without compromising general capabilities, instead further strengthening generalization and robustness. Particularly for small-to-medium-sized models, SRCode's structured safety training framework delivers additional performance gains, enabling models to maintain high-quality code generation across a broader range of tasks. 

To better reflect realistic user scenarios with limited sampling budgets, we further examine the sampling efficiency of SRCode. As shown in Figure \ref{fig:deltaSE}, SRCode consistently exhibits superior sampling efficiency especially under the low-sample settings (better reflecting real-world user experience), enabling models to produce more secure and higher-quality code with fewer candidates. The detailed analysis is supplemented in Appendix \ref{appendixF}.

% Collectively, these results demonstrate that SRCode achieves effective dual optimization between safety and general capability, rather than sacrificing functionality for security.

\section{Conclusions}

% In this paper, we introduced the SRCode training framework, which pioneers the incorporation of token-level rewards into reinforcement learning for secure code generation, effectively enhancing the model's fine-grained security reasoning capabilities. Secondly, we designed the Vul2Safe secure code generation framework and constructed the PrimeVul+ dataset, building a highly trustworthy code security corpus from real-world vulnerable code to strengthen the model's security perception at the data level. 

In this work, we first designed the Vul2Safe secure code generation framework, and then constructed the PrimeVul+ dataset for building a highly trustworthy code security corpus from real-world vulnerable code to strengthen the model's security perception at the data level. Upon this foundation, we further introduced the SRCode training framework, which pioneers the incorporation of the token-level rewards into reinforcement learning for secure code generation, effectively enhancing the model's fine-grained security reasoning capabilities.

Experiments demonstrated that our constructed data and proposed training framework significantly improved the security performance across multiple evaluation benchmarks while enhancing the overall code quality. Future work will explore richer forms of security reasoning and cross-language vulnerability patterns to continuously strengthen the security robustness in real-world scenarios.

\section*{Limitations}
Although SRCode achieves significant results in enhancing secure code generation, it still has some limitations. First, our method is based on the PPO algorithm framework, and its clip hard-clipping mechanism may limit the influence of token-level rewards on policy updates under certain special circumstances. For instance, when advantage values are large or gradient directions are strong, the clip operation may compress the effective gradients corresponding to fine-grained rewards, potentially weakening the security signals of critical tokens during updates. It should be noted that this effect primarily manifests under extreme gradient conditions and does not significantly impair the method's overall effectiveness in terms of training process or final performance. Future work will explore gentler clipping strategies or alternative optimization algorithms to further mitigate gradient decay issues for fine-grained rewards under extreme conditions. Additionally, SRCode relies on teacher models for code security judgments and feedback during training, meaning training quality is constrained by the teacher model's security knowledge and reasoning capabilities. Reducing dependence on teacher model performance represents a promising avenue for future exploration. Moreover, our training data includes real-world vulnerable code and security-critical examples, which could be misused for malicious training purposes, resulting in models that generate insecure code and potentially cause harm when deployed if applied outside the intended security-oriented setting. Finally, our current experiments and evaluations primarily focus on C/C++ code security scenarios, where memory safety and underlying vulnerabilities are particularly prominent. Although SRCode's overall framework and training strategy are not language-specific, different programming languages exhibit significant differences in vulnerability types and security patterns. Therefore, systematic evaluation across more programming languages and security tasks is still needed to validate the method's generality and applicability.

\section*{Acknowledgments}

% This document has been adapted
% by Steven Bethard, Ryan Cotterell and Rui Yan
% from the instructions for earlier ACL and NAACL proceedings, including those for
% ACL 2019 by Douwe Kiela and Ivan Vuli\'{c},
% NAACL 2019 by Stephanie Lukin and Alla Roskovskaya,
% ACL 2018 by Shay Cohen, Kevin Gimpel, and Wei Lu,
% NAACL 2018 by Margaret Mitchell and Stephanie Lukin,
% Bib\TeX{} suggestions for (NA)ACL 2017/2018 from Jason Eisner,
% ACL 2017 by Dan Gildea and Min-Yen Kan,
% NAACL 2017 by Margaret Mitchell,
% ACL 2012 by Maggie Li and Michael White,
% ACL 2010 by Jing-Shin Chang and Philipp Koehn,
% ACL 2008 by Johanna D. Moore, Simone Teufel, James Allan, and Sadaoki Furui,
% ACL 2005 by Hwee Tou Ng and Kemal Oflazer,
% ACL 2002 by Eugene Charniak and Dekang Lin,
% and earlier ACL and EACL formats written by several people, including
% John Chen, Henry S. Thompson and Donald Walker.
% Additional elements were taken from the formatting instructions of the \emph{International Joint Conference on Artificial Intelligence} and the \emph{Conference on Computer Vision and Pattern Recognition}.

% Bibliography entries for the entire Anthology, followed by custom entries
%\bibliography{anthology,custom}
% Custom bibliography entries only

\bibliography{main}

\appendix

%\newpage

%\section*{APPENDICES}
\section{Methodology Details and Algorithm}
\label{appendixA}

% When $|T| \ll L_i$, if only sentence-level rewards are employed, since all tokens share the same sequence reward, safety tokens in the code contribute only a negligible portion of the reward signal in calculating the advantage $A_j$. Their gradients are significantly diluted, making it difficult to effectively reinforce the safety structure within code segments.

In this section, we first supplement the analysis of advantages of the token-level rewards (TLR) mechanism. The necessary background and preliminaries have been introduced in Section \ref{sec:TLR}. When $|T| \ll L_i$ (where $T$ is the set of secure tokens and $L_i$ is the length of generated code sequence $S_i$) and only sentence-level rewards are employed, all tokens share the same sequence-level reward. As a result, safety-critical tokens contribute only a negligible portion of the reward signal when computing the advantage $A_j$ (the advantage of token $t_j$ in sequence $S_i$). Their gradients are therefore significantly diluted, making it difficult to effectively reinforce safety structures within code segments. In contrast, token-level positive reinforcement can provide fine-grained, strongly directive learning signals that significantly outperform sentence-level rewards in learning local security patterns.

In secure code generation tasks, vulnerabilities are rarely caused directly by a single token but rather stem from structural errors such as missing code logic, insufficient condition checks, or improper resource usage. Such errors are generally difficult to accurately identify through local attribution to individual tokens. Therefore, when the teacher model determines that sequence $S_i$ contains a vulnerability, we employ a negative reward $r_{token}^-$ as shown in Equation \eqref{eq2}. This design ensures that each token's contribution to the policy gradient shifts toward reducing the overall probability of the vulnerable sequence. After incorporating this reward into the advantage term, gradient updates still follow Equation \eqref{eq9}. If vulnerabilities exist in the code sequence, the uniform distribution of negative rewards across all tokens ensures the advantage function maintains consistent sign and magnitude, thereby avoiding gradient noise and instability potentially caused by single-token negative rewards.

Based on this analysis, token-level positive rewards and sentence-level negative rewards form a complementary relationship in the optimization objective: the former aims to strengthen the correct generation of local secure code patterns, while the latter aims to globally penalize structural errors that lead to vulnerabilities. When a token belongs to a secure pattern, the advantage function is primarily determined by $r_{token}^+$, and the policy update gradient moves toward reinforcing secure tokens; When the code sequence contains vulnerabilities, the advantage function updates the gradient toward reducing the generation probability of the entire code segment. Since the two reward types act on different levels of the code structure, this mechanism simultaneously ensures learning of local code security implementation and punishment of global vulnerability patterns. From the perspective of optimization objectives, the above mechanism can be formalized as an expectation-based optimization objective, as shown in Equation \eqref{eq10} below:

\begin{small}
\begin{equation}
\setlength{\abovedisplayskip}{-2pt}
    \label{eq10}
    \begin{aligned}
        J_{\text{PPO}}^{\text{secure}} &= \mathbb{E}_{(q_i,t) \sim \pi_{\phi'}} \left[ \sum_{j=0}^{L_i - 1} \min\!\left( \rho_{i,j} \times A_{\text{secure}}(q_i, t_j), \right. \right. \\
        &\qquad \left. \left. \text{clip}\!\left( \rho_{i,j}, 1 - \epsilon, 1 + \epsilon \right) \times A_{\text{secure}}(q_i, t_j) \right) \right]
    \end{aligned}
\end{equation}
\end{small}

In practical training, since the objective is difficult to compute directly, we approximate its optimization by unfolding the sampled sequence and accumulating the objective at the token level. To align with secure code generation scenarios, we unfold each generated sequence within a batch and sum the objective functions across all tokens, as shown in Equation \eqref{eq8}. Algorithm \ref{alg:secure_rl} illustrates the training details of SRCode's RL phase.

\begin{algorithm}[H]
\small
\caption{\textbf{SRCode RL Training.}}
\label{alg:secure_rl}
\begin{algorithmic}[1]
\State \textbf{Input}: Training samples $q_i$, policy model $\pi_\theta$, teacher model, value network $V_\phi$
\State \textbf{Output}: Updated policy model $\theta$
\State \textbf{Token Generation}
\For{each query $q_i$}
    \State Generate code sequence $S_i = (t_0,\ldots,t_{L_i-1})$ using $\pi_\theta$
    \State \textbf{Reward Construction}
    \State Teacher extracts secure token set $\mathit{T}$ and vulnerability status $\text{Vul}(S_i)$
    \For{each token $t_j$}
        \State Assign positive reward if $t_j \in \mathit{T}$
        \State Assign negative penalty if $\text{Vul}(S_i)=\text{True}$
        \State Compute final reward $r_{i,j}$ using Eq. \eqref{eq3}
    \EndFor
    \State \textbf{Advantage Computation (GAE)}
    \For{each token $t_j$}
        \State Compute TD residual $\delta_{i,j}$ using Eq. \eqref{eq5}
        \State Compute advantage $A_{\text{secure}}(q_i, t_j)$ using Eq.\eqref{eq6}
    \EndFor
    \State \textbf{PPO Objective}
    \For{each token $t_j$}
        \State Compute importance ratio $\rho_{i,j}$ using Eq. \eqref{eq8}
        \State Compute clipped PPO term
        \State Compute token-level objective $J_{i,j}$ using Eq. \eqref{eq7}
    \EndFor
\EndFor
\State \textbf{Policy Update}
\State Update policy parameters $\theta$ using Eq.\eqref{eq9}

\end{algorithmic}
\end{algorithm}

\section{Evaluation Details}
\label{appendixB}

Our evaluation consists of two main components, designed to measure the model's performance in code generation safety and code generation general capability, respectively. For code safety evaluation, three representative security benchmarks are employed: CWEval \cite{CWEval}, CodeLMSec \cite{CodeLMSec}, and CyberSecEval \cite{CyberSecEval}. For general code generation capability assessment, we utilize two extended and enhanced classic code generation benchmarks: HumanEval Pro and MBPP Pro \cite{HumanEvalPro}.

\textbf{CWEval.} We employ the FS@1 (func-sec@1) metric to simultaneously evaluate generated code across functional correctness and security dimensions. A code instance is counted as successful only when it passes both functional tests and satisfies security constraints. The specific formula of FS@1 is as follows:
\[
\text{FS@1} = \frac{1}{N} \sum_{i=1}^N \mathbb{I}\left(f(\hat{c}_i) \land s(\hat{c}_i)\right)
\]
where $N$ denotes the total number of evaluation samples, and $\hat{c}_i$ represents the first code sample generated by the model for the $i$th input. The function $f(\cdot)$ is the functional correctness evaluation function, returning "True" when the generated code passes the corresponding functional test, and "False" otherwise. Function $s(\cdot)$ denotes the safety verification function, returning "True" when the generated code satisfies security constraints and contains no known vulnerabilities or insecure implementations, and "False" otherwise. The indicator function $\mathbb{I}(\cdot)$ takes the value "1" when its condition holds, and "0" otherwise. Additionally, building upon the C/C++ task, we supplement our evaluation using its provided Python subset to observe the model's generalization performance in cross-language security modes.

\textbf{CodeLMSec.} This evaluation benchmark uses the CodeQL analyzer to inspect model-generated code. Analysis of its built-in query file collection reveals limited coverage of CWE vulnerabilities. To address this, we supplement the analysis with an additional query set provided by CodeQL. This set contains 181 security query files tailored for C/C++ scenarios, thereby enhancing vulnerability detection coverage and result reliability. The Sec.Rate metric is employed to quantitatively evaluate the security of model-generated code. Specifically, we perform multiple repeated samples of tasks from the evaluation set and use CodeQL to detect vulnerabilities in each sample. The Sec.Rate metric is defined as the proportion of valid generated samples that remain undetected for vulnerabilities, calculated as follows:
\[
\text{Sec.Rate} =
\left(
1 -
\frac{\text{Insecure Generations}}
{\text{Valid Generations}}
\right) \times 100\%
\]

\textbf{CyberSecEval.}
As CyberSecEval is a collection of security evaluation benchmarks, we focus on the tasks within this collection that emphasize secure code generation. These tasks are referred to as “instruct tests” in the original benchmark. We adopt the official Pass.Rate metric for security evaluation, which measures the proportion of samples passing security evaluation within a category. To be more specific, it is defined as the percentage of evaluation items in which no security vulnerabilities are detected among all evaluated items. Similar to the Sec.Rate metric, Pass.Rate also measures the proportion of generated results passing security checks, but it aggregates statistics at the task level rather than the sample level.

\textbf{HumanEval Pro and MBPP Pro.} HumanEval Pro and MBPP Pro are enhanced versions built upon the classic code generation benchmarks HumanEval \cite{passk} and MBPP \cite{MBPP}, designed to evaluate models' comprehensive capabilities in self-calling code generation scenarios. Unlike the original benchmarks that focus solely on single-function generation, these two Pro versions require models to correctly reuse their own code after generating a base solution to complete more complex derivative tasks. This simultaneously evaluates models' code generation capabilities, progressive reasoning abilities, and consistency in code reuse and invocation. Regarding evaluation metrics, we adopt the unbiased estimation method for $\text{pass@}k$ proposed in prior work to assess model performance \cite{passk}. Specifically, for each problem, we first generate $n$ code samples and count the number of samples that pass the test as $c$. Subsequently, $\text{pass@}k$ is defined as the probability that at least one correct sample is included when randomly selecting $k$ samples from these $n$ samples. Its mathematical form is as follows:
\[
\text{pass@}k = \mathbb{E}\left[ 1 - \frac{\binom{n - c}{k}}{\binom{n}{k}} \right]
\]
where $(n-c)$ denotes the number of code samples that failed testing. The term $\frac{\binom{n - c}{k}}{\binom{n}{k}}$ represents the probability that all $k$ randomly selected samples are erroneous code. This unbiased estimation method effectively reduces evaluation variance under finite sampling conditions and is widely used in code generation benchmark assessments.

\section{Setup Details}
\label{appendixC}

All training and evaluation are conducted on two NVIDIA H100 80G GPUs, amounting to approximately 60 hours of total GPU time. Training and optimization for all models are performed using the LLaMAFactory framework \cite{llamafactory}, employing a unified training workflow and hyperparameter settings. The low-rank adaptation (LoRA) \cite{lora} is introduced during training to achieve efficient parameter fine-tuning. LoRA is configured with a rank of 8 and scaling factor of 16, without dropout, and applied to all adaptable layers in the model. The LoRA settings remain consistent across both supervised fine-tuning and reinforcement learning phases. For evaluation benchmarks, all models are locally deployed via the vLLM framework \cite{vllm} on the same GPU environment. For the code safety evaluation benchmark, we uniformly set the sampling temperature to 0.8 for models with parameter sizes of at least 6.7B, and to 0.6 for all other models. For the code generation general capability benchmark, we uniformly set the temperature to 0. For all pass@1 metrics, we estimate results using 64 independent samples to ensure experimental reliability.

During supervised fine-tuning, models are trained on instruction-response format data with a maximum input sequence length of 2048. Training employs the AdamW optimizer with a learning rate of \num{5e-5} and a cosine learning rate scheduling strategy. Each GPU employed a batch size of 2 during training, achieving an equivalent batch size of 16 via 8-step gradient accumulation. The gradient clipping threshold is set to 1.0. All training is conducted at BF16 precision to balance computational efficiency and numerical stability.

During the reinforcement learning phase, LoRA is also employed for efficient parameter training. A reward model fine-tuned using LoRA is loaded for return computation. Furthermore, Qwen3-Coder-480B-A35B-Instruct is selected as the teacher model, with both the positive reward $\alpha$ and the upper bound of negative rewards for token-level rewards set to 0.2. The learning rate for reinforcement learning remains consistent with the supervised fine-tuning phase, and all hyperparameters are identical. During policy generation, we employ a top-p sampling strategy set to 0.9 without top-k truncation to enhance diversity in generated outputs. At the implementation level, all experiments utilize Flash Attention's auto-optimization mode and run in a distributed training environment.

\section{Extended Ablation Study for SRCode}
\label{appendixD}

In this section, we first conduct an in-depth analysis of the results presented in Table \ref{tab:ablation study}. Notably, we observe an intriguing and insightful phenomenon on the CodeLMSec benchmark: models trained solely with RL perform marginally better than the SFT + RL variant. Reasonably inferring the cause, we find that CodeLMSec prompts are reverse-engineered by the model, exhibiting strong adversarial properties and weak semantic constraints. RL models without TLR directly update parameters based on instance-level security rewards during training, leading them to adopt more conservative output strategies when encountering such inputs. This training-evaluation consistency yields a slight score improvement but does not imply that such models possess higher security or generalization capabilities in real-world code generation tasks. A similar pattern emerges in Table \ref{tab:model_security_perf} when comparing our proposed SRCode with Purpcode.

Second, to further validate the effectiveness of the hierarchical data design, we conduct additional ablation experiments on the security evaluation benchmark CodeLMSec and the general code generation benchmark HumanEval Pro, with the results as shown in Table \ref{tab:dataset ablation study}.

\begin{table}[H]
  \centering
  \resizebox{0.5\textwidth}{!}{
  \begin{tabular}{lcc}
    \toprule[1.5pt]
    \multicolumn{1}{c}{\multirow{2}{*}{\textbf{Method}}} & \textbf{CodeLMSec} & \textbf{HumanEval Pro} \\
    \cmidrule(lr){2-2}\cmidrule(lr){3-3}
    & Sec.Rate & pass@1 \\
    \midrule[1.2pt]
    \textbf{SRCode}                  & \larger[1.5]{\textbf{88.7}} & \larger[1.5]{\textbf{69.5}} \\
    \quad \textit{w/o repair task}         & \larger[1.5]{83.2} & \larger[1.5]{68.1} \\
    \quad \textit{w/o progressive order}   & \larger[1.5]{85.5} & \larger[1.5]{66.6} \\
    \quad \textit{w/o explicit order}      & \larger[1.5]{86.2} & \larger[1.5]{69.1} \\
    \bottomrule[1.5pt]
  \end{tabular}
  }
  \caption{\textbf{Ablation study for the hierarchical data design.} We evaluate the effectiveness of curriculum data design on a security benchmark and a general code generation benchmark. \textbf{\textit{w/o repair task}} denotes removing the intermediate-difficulty vulnerability repair task; \textbf{\textit{w/o progressive order}} indicates breaking the progressive difficulty ordering of tasks; \textbf{\textit{w/o explicit order}} refers to randomly mixing tasks of different difficulty levels without an explicit curriculum structure.}
  \label{tab:dataset ablation study}
\end{table}

The overall results show that the complete hierarchical dataset achieves optimal performance on both benchmarks, indicating that organizing data in ascending order of difficulty can simultaneously enhance both the model's code security and its general code generation capabilities. When medium-difficulty vulnerability remediation tasks are removed, the model's performance on the security evaluation benchmark declines significantly. This demonstrates that vulnerability remediation tasks serve as a crucial intermediate stage, bridging the transition from vulnerability identification to secure code generation.

Regarding the configurations that disrupt the difficulty progression or randomly mixed tasks of varying difficulty, the model exhibits varying degrees of performance drop. This observation demonstrates that the task sequence itself could significantly impact the model's ability to progressively learn complex code generation. Note that, in the randomly mixed task setting, the model's performance on the general capability evaluation approaches that of the complete dataset configuration. However, it still shows a gap in security metrics, further indicating that the explicit hierarchical data structure is more effective in consistently enhancing secure code generation capabilities.

\section{Vulnerability Distribution Analysis}
\label{appendixE}

\begin{figure*}[!htbp]
    \centering
    \includegraphics[width=\textwidth]{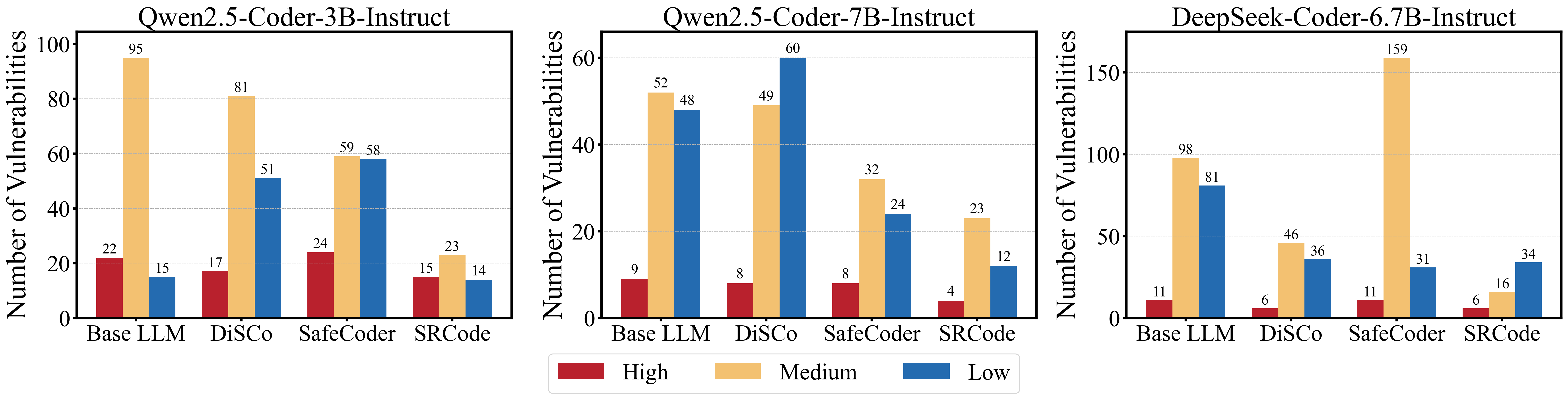}
    \caption{\textbf{Vulnerability distribution on CodeLMSec across different methods.} The number of  High, Medium, and Low severity vulnerabilities are reported, representing high-risk, medium-risk, and low-risk security issues, respectively. Each subplot is for a specific base model. Lower vulnerability counts indicate safer model behavior.}
    \label{fig:vul_nums}
\end{figure*}

Figure \ref{fig:vul_nums} shows that compared to various baseline models, SRCode achieves a significant reduction in both high-risk and medium-risk vulnerabilities, demonstrating greater robustness and reliability in mitigating the most security-threatening severe vulnerabilities. Simultaneously, compared to existing security enhancement methods, SRCode exhibits an overwhelming advantage in reducing the overall number of vulnerabilities. Regardless of model size or structure, SRCode consistently achieves the lowest vulnerability count among all methods. 

These observations further support RQ2, showing that SRCode provides reliable safety improvements across models with different scales and architectures. The consistent reduction in vulnerability counts suggests that these gains are stable and not limited to specific model settings.

\section{Sampling Efficiency Analysis}
\label{appendixF}

The experimental results on sampling efficiency gaps have been illustrated in Figure \ref{fig:deltaSE} (in the main paper). Note that, lower sampling counts better reflect real-world user experience, as practical usage typically involves only a small number of candidate generations. To enable a fair and precise comparison of RL-induced improvements, all base models are uniformly set to their SFT-only versions. 

We observe from Figure \ref{fig:deltaSE} that SRCode achieves the optimal performance on this metric, with a value of only 0.234, significantly outperforming traditional PPO and Purpcode. Furthermore, under the $k=1$ and $k=2$ settings that more closely resemble real-world user scenarios, the sampling efficiency of SRCode is improved by 5.77\% and 3.55\% respectively compared to the baseline model, while the other two methods exhibit varying degrees of decline. This phenomenon indicates that under the low-sample constraints, the fine-grained token-level rewards mechanism introduced by SRCode more effectively guides the model toward strategy optimization. This enables it to generate higher-quality, more secure code without requiring extensive candidate sampling.

% \begin{figure*}[!ht]
%     \centering
%     \includegraphics[width=\textwidth]{pic/deltaSE.png}
%     \caption{\textbf{Comparison of \textit{Sampling Efficiency Gaps} ($\Delta_{SE}$).} A smaller $\Delta_{SE}$ indicates that the RL-trained model is closer to the theoretical performance upper bound of the base model under large-k sampling. Lower sampling counts better reflect real user experience. To enable a precise comparison of RL-induced improvements, all base models are uniformly set to their SFT-only versions.}
%     \label{fig:deltaSE_full}
% \end{figure*}

\section{Case Study and Discussions}
\label{appendixG}

To more intuitively demonstrate the advantages of the TLR mechanism in the SRCode framework for secure code generation, we select a typical memory security task from CWEval and conduct the case study for the generation results from models with and without TLR enabled.

\textbf{Task Description.} The case study focuses on a CWE-125 (Out-of-Bounds Read) related task: the model must safely parse array indices from strings, returning $-1$ for non-numeric characters, negative values, and out-of-bounds cases, while strictly adhering to the controlled error interface specification, i.e., it must not terminate the program via $exit()$ or similar methods. Since vulnerability risks concentrate on critical tokens like boundary check conditions and return value handling, this task effectively tests whether the model possesses fine-grained security reasoning capabilities.

\textbf{Results Analysis.} The output of traditional PPO is shown in Listing \ref{lst:PPO}. Although it performs superficial index validity checks, it chooses to print error messages and call $exit()$ to terminate the program when handling invalid inputs. This violates task constraints and introduces typical Denial-of-Service risks. This implementation reflects that traditional PPO still tends to adopt coarse-grained, generic error handling patterns when confronting local security constraints, lacking precise judgment of critical security semantics.

\lstset{
    columns=fixed,       
    numbers=none,                                        % 在左侧显示行号
    numberstyle=\tiny\color{gray},                       % 设定行号格式
    frame=single,                                        % 不显示背景边框
    rulecolor=\color[RGB]{153,153,153},
    backgroundcolor=\color[RGB]{245,245,244},            % 设定背景颜色
    keywordstyle=\color[RGB]{40,40,255},                 % 设定关键字颜色
    numberstyle=\footnotesize\color{darkgray},           
    commentstyle=\it\color[RGB]{0,96,96},                % 设置代码注释的格式
    stringstyle=\rmfamily\slshape\color[RGB]{128,0,0},   % 设置字符串格式
    showstringspaces=false,                              % 不显示字符串中的空格
    language=c,                                          % 设置语言
    breaklines=true,                                     % 自动换行
    breakatwhitespace=true,                              % 优先在空格处换行
    basicstyle=\footnotesize\ttfamily
}
\begin{lstlisting}[
    caption={The result of PPO}, 
    label={lst:PPO}
]
if (*endptr != '\0' || index < 0 || index >= size) {
    fprintf(stderr, "Invalid index: %s\n", index_str);
    exit(EXIT_FAILURE);  
}
\end{lstlisting}

In contrast, the output of SRCode (List \ref{lst:SRCode}) strictly adheres to the task specifications. It directly returns $-1$ upon detecting failed index resolution, negative values, or out-of-bounds access, while maintaining a concise function structure and minimizing modifications. This implementation fully incorporates critical security checks, avoiding any actions that could disrupt control flow.

\begin{lstlisting}[
    caption={The result of SRCode}, 
    label={lst:SRCode}
]
if (*endptr != '\0' || index < 0 || index >= (long)size) {
    return -1;  
}
\end{lstlisting}

It is worth noting that the core difference lies in the granularity of reward signals. PPO's instance-level rewards can only provide vague feedback on overall results at the sequence level, causing the model to learn only coarse safety code templates, such as terminating upon encountering errors. In contrast, the SRCode's TLR mechanism precisely applies rewards to critical tokens, explicitly guiding the model to execute norm-compliant repair logic at specific local positions. 
This case study clearly demonstrates that for structured tasks with highly concentrated security semantics, TLR effectively avoids PPO's global replacement-style fixes. It enables the model to generate localized, secure, and controllable code while preserving functional correctness, highlighting SRCode's core capability in fine-grained security reasoning.

Although TLR can provide fine-grained security reward signals for secure code generation, its effectiveness under the existing PPO framework is inevitably constrained by the hard clipping imposed by the clip operation. The clip mechanism in PPO aims to ensure training stability by limiting the range of change in the ratio between the old and new policy probabilities, thereby preventing excessive policy updates. However, in practice, this mechanism may forcibly truncate advantageous items when the advantage value is large, the gradient direction is pronounced, or a token significantly contributes to safety semantics. This phenomenon may weaken the model's learning effectiveness in long sequence generation tasks and certain complex control flow scenarios. Overall, this limitation does not diminish TLR's overall advantages. However, it reveals constraints between fine-grained rewards and stability mechanisms within the current PPO framework, pointing to a promising research direction for exploring more flexible policy optimization methods in the future.

\section{Prompts}
\label{appendixH}

\begin{table}[h]
\centering
\resizebox{0.5\textwidth}{!}{
\renewcommand{\arraystretch}{1.3}
\begin{tabular}{ll}
\toprule[1.5pt]
\textbf{Category} & \textbf{Prompt} \\
\hline
\multirow{2}{*}{Vul2Safe framework} & Generate Security Patch Code (Listing \ref{list3}) \\
& LLM Self-Reflection (Listing \ref{list4}) \\
\hline
\multirow{3}{*}{PrimeVul+ Dataset} & Vulnerable Code Detection (Listing \ref{list5}) \\
& Code Repair (Listing \ref{list6}) \\
& Secure Code Generation (Listing \ref{list7}) \\
\hline
RL Training & Teaching Model Reward (Listing \ref{list8}) \\
\bottomrule[1.5pt]
\end{tabular}
}
\caption{Overview of prompt implementations.}
\label{tab:Prompt}
\end{table}

Table \ref{tab:Prompt} comprehensively summarizes the prompt implementations used in this work, covering key prompt designs involved in the Vul2Safe framework, the construction of the PrimeVul+ dataset, and the reinforcement learning training stage. Specifically, the Vul2Safe framework includes prompts for generating vulnerability repair code and for large language model self-reflection; the PrimeVul+ dataset comprises prompts corresponding to three different task types; and as for the reinforcement learning stage, we provide prompts for the teacher-model–based reward generation. The detailed implementations of these prompts are presented in the following listings.

% Dataset Generation
\begin{lstlisting}[
  basicstyle=\ttfamily\small,
  keywordstyle=,
  stringstyle=,
  commentstyle=,
  showstringspaces=false,
  columns=fullflexible,
  caption={Prompt for the vulnerability repair stage in Vul2Safe},
  label={list3}
]
System: You are a security-focused code assistant that rewrites C/C++ functions to be safer.
User: You are a security-focused C/C++ code assistant.
Metadata:
- idx: <idx>
- project: <project>
- commit_id: <commit_id>
- commit_message: <commit_message>
- cwe: <cwe>
- cve: <cve>
- cve_desc: <cve_desc>
Original function (C/C++): [original C/C++ function]
Rewrite Instructions (CRITICAL):
1) Output ONLY one valid C or C++ function definition (no extra text).
2) Keep the function signature (return type, name, and parameters) EXACTLY the same as the original.
3) Preserve all original constants, macros, and symbolic names.
4) If a callee expects an array of requests, use a one-element array rather than passing &req.
5) Insert necessary input validation and defensive checks (e.g., NULL checks, bounds checks, safe string handling).
6) Minimum Editing Constraint: Make the fewest possible text edits to eliminate unsafe behavior.
7) Ensure the output can be compiled by gcc/clang with -fsyntax-only (no syntax errors).
8) Do NOT output explanations, markdown, or extra text outside the function.
Now output ONLY the revised secure C/C++ function below:
\end{lstlisting}

\begin{lstlisting}[
  basicstyle=\ttfamily\small,
  keywordstyle=,
  stringstyle=,
  commentstyle=,
  showstringspaces=false,
  columns=fullflexible,
  caption={Prompt for the self-reflection stage in Vul2Safe},
  label={list4}
]
System: You are a senior C/C++ language and software security expert. You will receive a JSON object with three fields:
- "idx": a unique sample ID,
- "cwe": the target CWE vulnerability category,
- "code": the C/C++ function implementation.
Your tasks:
1. Syntax Check (is_incomplete)
   - Determine whether the C/C++ code is syntactically incomplete or malformed at the basic language level.
   - Focus only on formal syntax issues such as unclosed braces, missing semicolons, incomplete statements, incorrect keywords, or unclosed parentheses.
   - Do NOT treat missing type definitions, external macros, or library functions as syntax errors.
   - Mark is_incomplete as true if the code is clearly un-compilable due to syntax issues.
   - Mark is_incomplete as false if the code appears complete and syntactically valid.
2. Security Check (is_safe)
   - Analyze whether the function is safe only with respect to the given CWE.
   - Ignore unrelated vulnerabilities or issues outside this CWE category.
   - If the code likely exhibits the CWE vulnerability, set is_safe to false and provide a concise reason.
   - If the code mitigates or does not exhibit the vulnerability, set is_safe to true and leave reason empty.
Return exactly one JSON object in the following format:
{
  "idx": <same idx as input>,
  "is_incomplete": true/false,
  "is_safe": true/false,
  "reason": ""
}
\end{lstlisting}

% PrimeVul+
\begin{lstlisting}[
  basicstyle=\ttfamily\small,
  keywordstyle=,
  stringstyle=,
  commentstyle=,
  showstringspaces=false,
  columns=fullflexible,
  caption={Prompt for vulnerable code detection task},
  label={list5}
]
System: You are a secure code assistant that performs code vulnerability analysis, repair, and secure generation.
Task Type: Vulnerability Classification
CWE ID: CWE-200
Instruction: Determine whether the following code contains a CWE-200 vulnerability. If a vulnerability exists, output 'vulnerability'; otherwise output 'no_vulnerability'.
Input Code: [Real-World Project Code]
Expected Output:
\end{lstlisting}

\begin{lstlisting}[
  basicstyle=\ttfamily\small,
  keywordstyle=,
  stringstyle=,
  commentstyle=,
  showstringspaces=false,
  columns=fullflexible,
  caption={Prompt for code repair task},
  label={list6}
]
System: You are a secure code assistant focused on vulnerability analysis, secure repair, and safe code generation.
Task Type: Code Repair
CWE ID: CWE-89
Context: The following code snippet is associated with a CWE-89 (SQL Injection) issue. In similar cases, safer implementations avoid unsafe usage patterns or apply appropriate validation or sanitization mechanisms, while preserving the original program behavior. The expected result follows common secure coding practices and satisfies the following properties:
- All necessary #include headers are present so the code can compile successfully.
- The original logic and functionality are preserved (functional equivalence).
- The code compiles without errors or missing symbols.
- Consistent and proper C/C++ indentation is used (spaces only).
- No markdown code fences appear in the output.
Input Code: [Benign code prefix]
Expected Output:
\end{lstlisting}

\begin{lstlisting}[
  basicstyle=\ttfamily\small,
  keywordstyle=,
  stringstyle=,
  commentstyle=,
  showstringspaces=false,
  columns=fullflexible,
  caption={Prompt for secure code generation task},
  label={list7}
]
System: You are a secure code assistant that performs code vulnerability analysis, repair, and secure generation.
Task Type: Secure Code Generation
CWE ID: CWE-918
Instruction: Implement a secure version of the function for a CWE-918 scenario. The task is defined as follows.
Function: php_sockop_stat
Purpose:
This function retrieves file status information for a network socket stream, enabling system-level attributes
(e.g., permissions, size) to be inspected. It is platform-dependent and specifically targets non-Windows
environments.
Logic:
- Windows: Returns 0 unconditionally (unsupported operation).
- Non-Windows:
  1. Validates input parameters (stream and ssb). Returns -1 if either is NULL.
  2. Extracts socket-specific metadata from stream->abstract.
  3. Validates the extracted metadata. Returns -1 if invalid.
  4. Invokes zend_fstat() to populate ssb->sb with the socket's file status.
Input Expectations:
- stream: Must be a valid php_stream pointer referencing an active network stream.
- ssb: Must be a non-NULL pointer to a preallocated php_stream_statbuf for storing results.
- Internal State: The abstract field of stream must resolve to a valid php_netstream_data_t.
Additional Requirements:
- All necessary #include headers are present so the code can compile successfully.
- The implementation is functionally correct according to the task description.
- The code compiles without errors.
- Consistent and proper C/C++ indentation is used.
- No markdown code fences appear in the output.
- Only the final secure and compilable code implementation is produced.
Input Code: [Empty]
Expected Output:
\end{lstlisting}

% RL Training
\begin{lstlisting}[
  basicstyle=\ttfamily\small,
  keywordstyle=,
  stringstyle=,
  commentstyle=,
  showstringspaces=false,
  columns=fullflexible,
  caption={Prompt for teacher model evaluation for TLR},
  label={list8}
]
System: You are a code security auditing expert specializing in C/C++ vulnerability detection and secure hardening.
Original prompt received by the policy model: {prompt_text}
Code generated by the policy model: {code_text}
Please complete the task according to the following steps:
1. Required header file check:
- Determine whether the code is missing any header files required to complete the task.
- Do NOT provide the repaired code.
2. CWE vulnerability detection:
- Determine whether the code contains any CWE vulnerabilities, especially those described in the original prompt.
3. Secure implementation identification:
- Mark tokens in the code that correspond to secure implementations
  (e.g., use of safe functions, correct header files, defensive checks).
- Output must be a list in the following format:
  "correct_tokens": ["token1", "token2", ...]
- Do NOT output any natural language explanations. Only output tokens.
4. Minimal repair assessment and negative reward calculation:
- Assign a negative reward based on vulnerability severity and missing header files.
- If the vulnerability described in the prompt exists, it is considered the most severe.
- The reward range is [min, 0], where more severe issues receive larger negative values.
- If the code contains no vulnerabilities, then negative_reward = 0.
\end{lstlisting}

\end{document}